\documentclass[aps,prl,reprint,groupedaddress,showpacs,amsmath,amssymb,floatfix]{revtex4-1}

\usepackage{graphicx}
\usepackage{bm}
\usepackage{color}
\usepackage[normalem]{ulem}
\usepackage{cleveref}
\usepackage{amsmath}
\usepackage{braket}
\usepackage{xspace}

\newcommand{\wse}{WSe$_2$\xspace}
\newcommand{\rucl}{$\alpha$-RuCl$_3$\xspace}
\newcommand{\sio}{SiO$_2$\xspace}

\newcommand{\ko}{k$\Omega$\xspace}
\newcommand{\um}{$\mu$m\xspace}
\newcommand{\lldens}{$\times10^{10}$ cm$^{-2}$\xspace}
\newcommand{\ldens}{$\times10^{11}$ cm$^{-2}$\xspace}
\newcommand{\dens}{$\times10^{12}$ cm$^{-2}$\xspace}
\newcommand{\hdens}{$\times10^{13}$ cm$^{-2}$\xspace}
\newcommand{\cmvs}{cm$^{2}$/Vs\xspace}

\begin{document}

\title{Charge-transfer Contact to a High-Mobility Monolayer Semiconductor}

\author{J. Pack$^{1}$}
\author{Y. Guo$^{1}$}
\author{Z. Liu$^{1}$}
\author{B.S. Jessen$^{1}$}
\author{L. Holtzman$^{2}$}
\author{S. Liu$^{3}$}
\author{M. Cothrine$^{4}$}
\author{K. Watanabe$^{5}$}
\author{T. Taniguchi$^{6}$}
\author{D.G. Mandrus$^{4}$$^{7}$}
\author{K. Barmak$^{2}$}
\author{J. Hone$^{3}$}
\author{C.R. Dean$^{1}$$^{\dag}$}

\affiliation{$^{1}$Department of Physics, Columbia University, New York, NY 10027, USA}
\affiliation{$^{2}$Department of Applied Physics and Applied Mathematics, Columbia University, New York, New York 10027, United States}
\affiliation{$^{3}$Department of Mechanical Engineering, Columbia University, New York, NY 10027, USA}
\affiliation{$^{4}$Department of Materials Science and Engineering, University of Tennessee, Knoxville, Tennessee 37996, United States}
\affiliation{$^{5}$Research Center for Electronic and Optical Materials, National Institute for Materials Science, 1-1 Namiki, Tsukuba 305-0044, Japan}
\affiliation{$^{6}$Research Center for Materials Nanoarchitectonics, National Institute for Materials Science,  1-1 Namiki, Tsukuba 305-0044, Japan}
\affiliation{$^{7}$Materials Science and Technology Division, Oak Ridge National Laboratory, Oak Ridge, Tennessee 37831, United States}

\date{\today}

\maketitle

\textbf{Two-dimensional (2D) semiconductors, such as the transition metal dichalcogenides, have demonstrated tremendous promise for the development of highly tunable quantum devices. Realizing this potential 
requires low-resistance electrical contacts that perform well at low temperatures and low densities where quantum properties are relevant. Here we present a new device architecture for 2D semiconductors that utilizes a charge-transfer layer to achieve large hole doping in the contact region, and implement this technique to measure magneto-transport properties of high-purity monolayer WSe$\mathbf{_2}$. We measure a record-high hole mobility of 80,000~\cmvs and access channel carrier densities as low as $\mathbf{1.6\times10^{11}\  cm^{-2}}$, an order of magnitude lower than previously achievable. Our ability to realize transparent contact to high-mobility devices at low density enables transport measurement of correlation-driven quantum phases including observation of a low temperature metal-insulator transition in a density and temperature regime where Wigner crystal formation is expected, and observation of the fractional quantum Hall effect under large magnetic fields.  The charge transfer contact scheme paves the way for discovery and manipulation of new quantum phenomena in 2D semiconductors and their heterostructures.}

Semiconducting van der Waals materials provide a rich and versatile platform to study quantum many-body ground states. The ability to isolate single monolayer flakes allows direct access to the electron gas, enabling wide-ranging opportunities to both interrogate and manipulate the electronic states\cite{wilsonExcitonsEmergentQuantum2021,makSemiconductorMoireMaterials2022}. Further opportunities for band engineering arise through spatially periodic electrostatic gating, lithographic patterning, and interfacing with lattice-mismatched or rotated materials to form hetero- and homobilayers \cite{makSemiconductorMoireMaterials2022,shiGatetunableFlatBands2019}. The transition metal dichalcogendides (TMDs) are the most widely studied van der Waals semiconductors due to the material cleanliness and unique optoelectronic properties. 
At low carrier densities, electron-electron interactions play a dominant role, since the large effective mass suppresses the kinetic energy relative to the Coulomb energy\cite{larentisLargeEffectiveMass2018}.  A wide range of interaction-induced phenomena have been observed in TMDs,
including fractional quantum Hall states\cite{shiOddEvendenominatorFractional2020}, integer and fractional Chern insulators\cite{liQuantumAnomalousHall2021,fouttyMappingTwisttunedMultiband2023,caiSignaturesFractionalQuantum2023,zengThermodynamicEvidenceFractional2023,parkObservationFractionallyQuantized2023,xuObservationIntegerFractional2023}, and exciton insulators \cite{maStronglyCorrelatedExcitonic2021,nguyenPerfectCoulombDrag2023,qiPerfectCoulombDrag2023}.  However, only a small minority of these studies have employed electrical transport measurements, owing to the difficulty of making Ohmic contacts that remain transparent at low temperature and at low carrier density\cite{allainElectricalContactsTwodimensional2015}. 

Making electrical contacts to monolayer TMDs requires overcoming the well-known challenges intrinsic to making Ohmic contact to any semiconductor, such as Schottky barrier formation and Fermi level pinning\cite{allainElectricalContactsTwodimensional2015,wangMakingCleanElectrical2022}. Work-function tuned metals can help to reduce the Schottky barrier height, whereas doping the semiconductor in the contact region can be used to reduce the barrier width. However, techniques typically used to realize transparent electrical contact to bulk semiconductors, such as ion implantation and diffused metal contacts, cannot be applied in monolayer TMDs due to their atomically thin structure\cite{wangMakingCleanElectrical2022}. To overcome these limitations, contacts made from transferred metals\cite{liuApproachingSchottkyMott2018,movvaHighMobilityHolesDualGated2015,jungTransferredContactsPlatform2019}, van der Waals materials\cite{xuUniversalLowtemperatureOhmic2016}, and semimetals\cite{shenUltralowContactResistance2021} have been applied to TMDs. Other approaches have used surface treatments to fabricate heavily doped contacts\cite{borahLowResistancePTypeOhmic2021} or employed selective etches to make edge contacts\cite{caiBridgingGapAtomically2022}. While these techniques seek to reduce the effective barrier at the contacts, no approach has demonstrated high performance contact to high-mobility monolayer semiconductors at the low temperatures and low densities where electron-electron interactions significantly modify electronic properties.

 \begin{figure*}
\includegraphics[width=180mm]{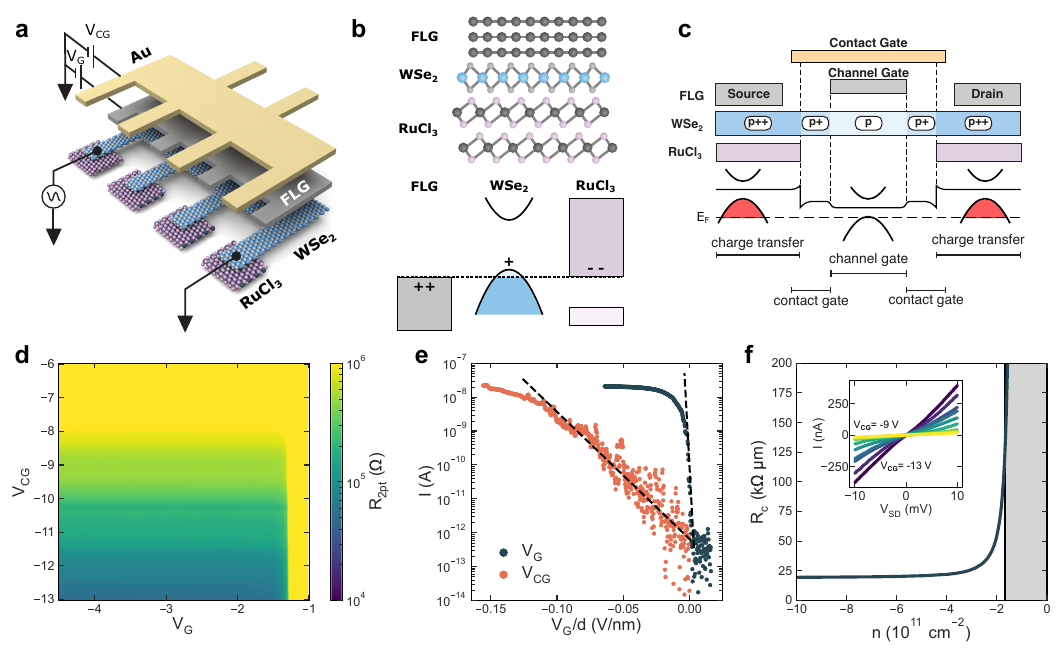}
\caption{{\bf{Charge-Transfer Contact Architecture.}}  \textbf{a}, Schematic illustration of a \wse device integrating charge-transfer contacts. Contact metal is omitted from the diagram to show the \wse - \rucl interface. \textbf{b}, Band diagram illustrating the expected charge redistribution when graphite, \wse, and \rucl are brought in contact. The modification of the \wse work function serves to reduce the Schottky barrier. \textbf{c}, Diagram of a \wse device with charge-transfer contacts highlighting the doping profile and the band alignment between the contacts and channel. Fermi-level pinning is identified at the \rucl edge which limits the contact resistance. \textbf{d}, Two terminal resistance at 1.5 K as a function of $V_{CG}$ and $V_G$. \textbf{e}, Comparison of the device response to $V_{G}$ and $V_{CG}$ at 300 mK. Gate voltages are normalized to the corresponding BN thickness $d$. \textbf{f}, Contact resistance at 300 mK as a function of channel density which remains low over a wide range of density. Shaded region emphasizes densities below 1.7\ldens where the contact resistance exceeds 100 \ko\um. Inset shows I-V characteristics of the device varying the contact gate between -9 V and -13 V in steps of 0.5 V. At the maximum $|V_{CG}|$, the device shows a linear I-V response.}
\end{figure*} 

Here we demonstrate a novel contact scheme that utilizes a van der Waals electron acceptor, \rucl\cite{mashhadiSpinSplitBandHybridization2019,rizzoChargeTransferPlasmonPolaritons2020,wangModulationDopingTwoDimensional2020}, to achieve highly-transparent p-type contact to \wse.  We interface the \rucl to one side of the \wse to induce large hole-doping in the contact region, with few layer graphite on the opposite side forming a metallic contact. Introducing both the charge transfer and contact layers to the 2D semiconductor by mechanical assembly allows us to realize a fully contacted and encapsulated 2D semiconductor in an all-van der Waals heterostructure.   
Charge transfer contact to flux-grown  \wse\cite{liuTwoStepFluxSynthesis2023}, allows us to study a higher mobility 2D semiconductor than all past monolayer and bilayer TMD devices, and with high performance contacts that persist to an order of magnitude lower density than those reported in past generations of devices. 
This allows us to study quantum transport properties at low carrier densities and under a strong magnetic field where interactions between carriers become prominent. We identify a low-density metal-to-insulator transition in a dilute regime where a Wigner crystal is expected to be stabilized. We also report the first transport signatures of the fractional quantum Hall effect (FQHE) in monolayer \wse.

\begin{figure*}
\includegraphics[width=110mm]{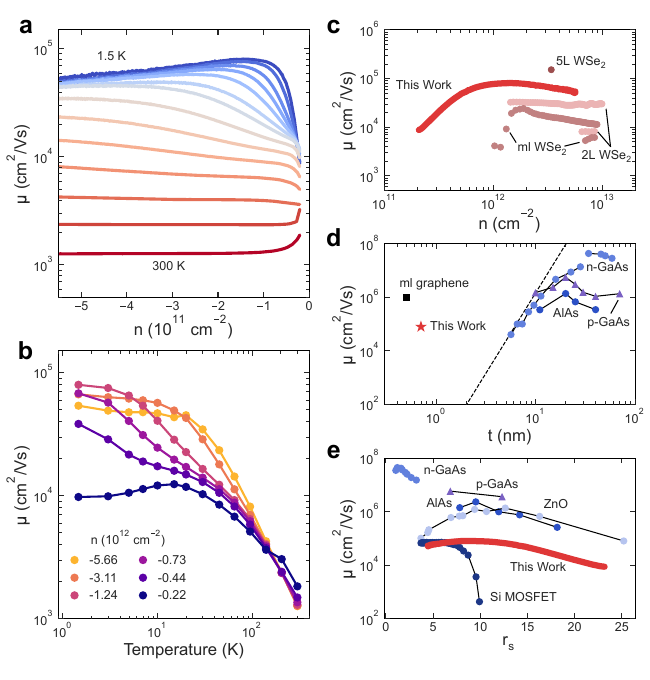}
\caption{{\bf{Transport Properties of Low Density WSe$\mathbf{_2}$.}} \textbf{a}, Hall mobility as a function of density from room temperature to 1.5 K. \textbf{b}, Hall mobility from room temperature to 1.5 K at multiple channel densities. \textbf{c}, Comparison of this work with past measurements of low-temperature Hall mobility as a function of carrier density in monolayer and few-layer \wse \protect\cite{shihSpinselectiveMagnetoconductivityWSe2023,joeTransportStudyCharge2023,caiBridgingGapAtomically2022,movvaMagnetotransportStudiesTungsten2018}. Using charge-transfer contacts, higher mobilities are accessible at lower densities. \textbf{d}, Mobility as a function of well thickness showing that the mobility realized in monolayer \wse exceeds what is possible in traditional quantum wells\protect\cite{chungUltrahighqualityTwodimensionalElectron2021,chungRecordqualityGaAsTwodimensional2022,chungMultivalleyTwodimensionalElectron2018,kamburovInterplayQuantumWell2016} at an equivalent thickness. Dashed line shows the expected $t^6$ scaling of mobility due to interfacial roughness in conventional 2DEGs. The value of mobility shown for graphene is a lower bound. \textbf{e}, Comparison of mobility as a function of $r_s$ for this work and other semiconducting 2DEGs\protect\cite{chungUltrahighqualityTwodimensionalElectron2021,chungRecordqualityGaAsTwodimensional2022,falsonReviewQuantumHall2018,kravchenkoPossibleMetalinsulatorTransition1994,chungMultivalleyTwodimensionalElectron2018}}
\end{figure*} 

\begin{figure}
\includegraphics[width=80mm]{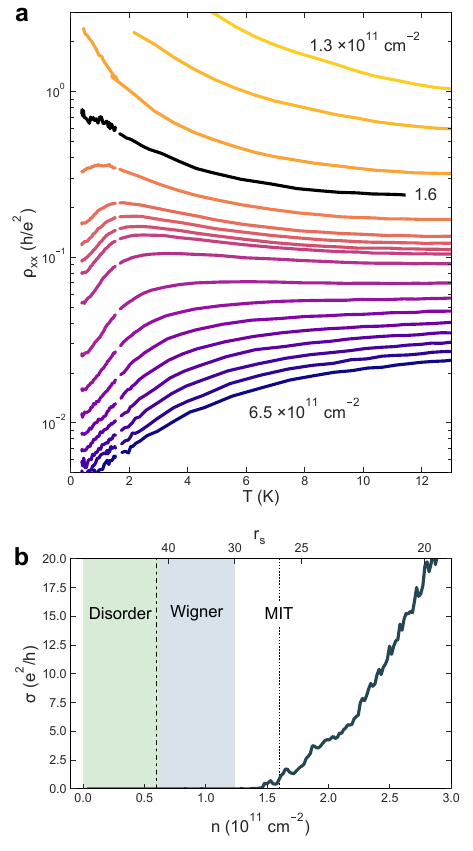}
\caption{{\bf{Low-Density Metal to Insulator Transition.}} \textbf{a}, Resistivity below 30 K for densities between 1.3\ldens and 6.5\ldens showing a metal to insulator transition around 1.6\ldens. \textbf{b}, Conductivity at 300 mK as a function of channel density. Shaded regions indicate the range of densities where a Wigner crystal is expected to be stabilized (blue) and where Anderson localization is expected (green). The critical density for the metal to insulator transition is shown as a dashed line. A dashed line at low density marks the density of charged defects identified in STM measurements of similar crystals \cite{liuTwoStepFluxSynthesis2023}.}
\end{figure} 

\begin{figure*}
\includegraphics[width=162mm]{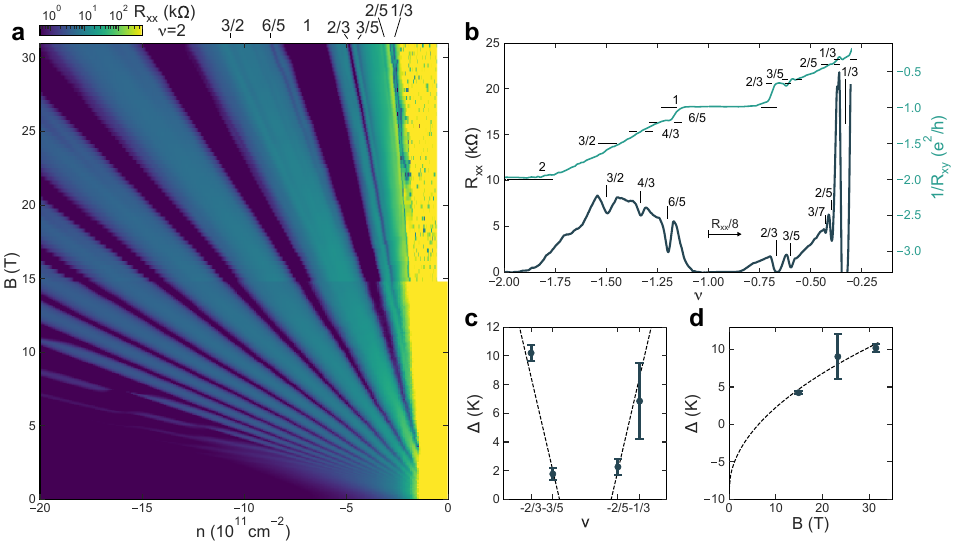}
\caption{{\bf{Fractional Quantum Hall Effect.}} \textbf{a}, $R_{xx}$ as a function of magnetic field and density below 2\dens at 300 mK. Data below and above 15 T were measured in different thermal cycles. \textbf{b}, $R_{xx}$ and $1/R_{xy}$ measured at 31.4 T for $\nu>-2$. $R_{xx}$ is scaled for $-1<\nu<0$. $R_{xx}$ displays minima or zeros at the labelled filling fractions. \textbf{c}, Activation gaps for the observed fractional quantum Hall states in the N=0 level as a function of filling fraction. A linear trend is shown following the expected scaling of the gap for Laughlin states. \textbf{d}, Magnetic field dependence of the measured activation gap at $\nu=2/3$. A $\sqrt{B}$ fit is shown as a trend in the data. The intercept at zero magnetic field corresponds to the disorder broadening $\Gamma$.}
\end{figure*} 

Figure 1a shows a cartoon schematic of the charge-transfer contact architecture, in which \rucl is incorporated so as to dope only the contact region of a h-BN encapsulated monolayer \wse device. As illustrated in Fig. 1b, the charge-transfer layer modifies the band alignment between the \wse and the contact metal (few-layer graphene in this case), reducing the Schottky barrier\cite{choModulationDopingSingleLayer2022}. Measurements of separate Hall bar structures in which large area \rucl uniformly dopes both the contact and channel regions (see SI) indicate that coupling to \rucl induces a hole density of 3.25\hdens in the \wse layer, with a contact resistance as low as 1.7 \ko\um, consistent with this microscopic understanding. 

Figure 1c shows a detailed cross-section diagram of the device structure with \rucl integrated in the contact region only. The device  profile consists of three doping regions.  In the vicinity of the metal contact, \rucl induces high hole doping.  In the channel region, the carrier density is controlled by a channel gate, capable of varying the hole density from fully depleted at the band edge to approximately 5\dens.  Between these two regions, a contact gate sets an intermediate hole density, thereby smoothing the transition between the contact and channel regions. A detailed description of the fabrication process we use to achieve this structure is given in the Methods section.

Fig. 1d plots the two-terminal device resistance as a function of channel gate voltage ($V_G$) and contact gate voltage ($V_{CG}$) at 1.5 K. The resistance increases sharply at the threshold $V_G$ where carriers are depleted from the channel. In contrast, the two terminal resistance changes much more slowly with varying $V_{CG}$.
The asymmetry between the sharpness of the change in resistance for the top gate and contact gate is reflected in the subthreshold slope, which is 1.2 V/dec for the contact gate compared to 7 mV/dec for the channel gate (see SI). When normalized by the dielectric thickness (Fig. 1e), the subthreshold slope of the contact gate still exceeds that of the channel gate by nearly a factor of 20. We interpret the comparatively high subthreshold slope of the contact gate response to be the result of Fermi-level pinning at the lateral boundary where the \rucl terminates.  The primary effect of the gate contact is to mitigate the resulting band-bending, illustrated in the schematic in Fig. 1c\cite{haratipourFundamentalLimitsSubthreshold2016,wangTransferredMetalGate2021}. 

Fig. 1f shows the contact resistance at 300 mK measured by subtracting the bulk resistance contribution from the two-terminal resistance, with -13 V applied to the contact gate.  The contact resistance measures approximately 20 \ko\um at high density and remains nearly constant down to 4\ldens before sharply increasing, rising to approximately 100 \ko\um at a density of 1.7\ldens. Fig. 1f inset shows two terminal IV response for varying voltage applied to the contact gate.   For large bias we observe linear response, consistent with Ohmic contact. 
We note that the contact resistance is larger than achieved when \rucl uniformly dopes both the contact and channel regions. For local contact doping, we believe the resistance is limited by the ability for the contact gate to populate carriers at the \rucl boundary. Further improvements in contact resistance could come from improved gate efficiency or interfacial engineering to soften the doping profile at this junction \cite{liSemiconductorPhysicalElectronics2006}.

Figure 2a shows the Hall mobility as a function of carrier density from room temperature down to cryogenic temperatures. At 300 K, the mobility exceeds 1000~$cm^{2}/Vs$ and remains constant over a wide range of density. Below 20 K, the mobility varies non-monotonically with carrier density: it first increases with increasing density below 1\dens and then decreases at higher densities. In high mobility 2DEGs, non-monotonic density dependence has been attributed to the presence of multiple scattering mechanisms relevant in different ranges of density; often, scattering from charged impurities is dominant at low densities and other mechanisms are dominant at high density \cite{huangScatteringMechanismsStateoftheart2022}. Recent transport studies of monolayer \wse have attributed the decreasing mobility with increasing density to short-range scattering from Se vacancies in the \wse monolayer\cite{joeTransportStudyCharge2023}. Figure 2b shows the temperature dependence of the mobility at multiple channel densities. At high temperatures, the mobility follows a power law as expected for scattering from optical phonons\cite{maChargeScatteringMobility2014,movvaHighMobilityHolesDualGated2015}. The value of the exponent varies with density between 0.8-1.6 which agrees with previous measurements of phonon-limited mobility in few-layer and monolayer \wse\cite{movvaHighMobilityHolesDualGated2015,joeTransportStudyCharge2023}. The mobility saturates at low temperatures, likely due to scattering from local or remote charged impurities.

Figure 2c compares the low-temperature Hall mobility as a function of carrier density measured in this work with past measurements of monolayer and few layer \wse\cite{shihSpinselectiveMagnetoconductivityWSe2023,movvaMagnetotransportStudiesTungsten2018,caiBridgingGapAtomically2022,joeTransportStudyCharge2023}. 
Using monolayers derived from \wse crystals grown with an optimized flux synthesis method \cite{liuTwoStepFluxSynthesis2023} we measure mobilities as high as 80,000 \cmvs and a mobility edge as low as 1.5\ldens at low temperatures. This is an improvement over past monolayer devices using different contact schemes and bulk crystal qualities which have achieved a maximum mobility of 25,000 \cmvs and reported transport response  for densities above 1.6\dens \cite{joeTransportStudyCharge2023}.

Fig. 2d shows a comparison plot of mobility vs quantum well or layer thickness, $t$, for a variety of 2DEG systems. In confined quantum wells, the mobility is strongly affected by scattering from interfacial roughness, which causes mobility to scale as $t^6$\cite{maChargeScatteringMobility2014,kamburovInterplayQuantumWell2016}. Due to the absence of interfacial roughness,  \wse demonstrates a unique scaling advantage, with the mobility far exceeding the $t^6$ limit for conventional 2DEGs. This suggests that the mobility measured in our device is limited by internal or external charged impurities and substantial mobility enhancement in the monolayer limit may still be possible by further refinements in \wse synthesis or heterostructure assembly\cite{rhodesDisorderVanWaals2019,liuTwoStepFluxSynthesis2023}.

The combination of large effective mass, high mobility, extreme confinement and low carrier densities make monolayer \wse a particularly promising platform to host correlated many body ground states. The expected influence of the Coulomb interaction in determining the electronic properties of a 2DEG is often characterized by the parameter $r_s=m^{*}/(m_{0}a_{B}\kappa\sqrt{\pi n})$, where $m^{*}/m_0$ is the band mass relative to the bare electron mass, $\kappa$ is the effective dielectric constant, and $a_B$ is the Bohr radius. Owing to the relatively large band mass (0.45 $m_0$)\cite{fallahazadShubnikovHaasOscillations2016} and low $\kappa$ (4.5)\cite{larentisLargeEffectiveMass2018}, $r_s$ is larger in \wse than many other 2D semiconductors at the same density. Fig. 2e compares the mobility as a function of $r_s$ between this work and other well-studied 2D electron systems \cite{chungUltrahighqualityTwodimensionalElectron2021,chungRecordqualityGaAsTwodimensional2022,falsonReviewQuantumHall2018,kravchenkoPossibleMetalinsulatorTransition1994,chungMultivalleyTwodimensionalElectron2018}. We find that monolayer \wse displays high mobility over a large range of $r_s$, similar to  AlAs and ZnO devices, and exceeding the range of $r_s$ values accessible in Si MOSFET devices by more than a factor of 2. 

The low sample disorder together with the ability to maintain high transparency contact to low electron densities allows us to map the low temperature metal to insulator transition (MIT). Fig. 3a shows the resistivity as a function of temperature at densities between 1.3\ldens and 6.5\ldens where we observe a MIT. The critical density (defined as the density where d$\rho$/d$T$ changes sign in the low-temperature limit) is approximately 1.6\ldens ($r_s=26.4$). 

The critical density is near the value of $r_s$ where a MIT has been observed in ZnO quantum wells and attributed to the formation of a Wigner crystal\cite{falsonCompetingCorrelatedStates2022}. Fig 3b plots the conductivity of the sample at $T=300$ mK as a function of density and highlights the expected mechanisms for insulating behavior at these low densities. We estimate the range of density that we might expect disorder-induced localization using the Ioffe-Regel-Mott criterion, $\Gamma=E_F$ \cite{ahnDensitytunedEffectiveMetalinsulator2023}. We calculate $\Gamma=4$ K from Shubnikov-de Haas oscillations (see SI) and find that the density where the IRM criterion is satisfied coincides with the charged impurity density measured by STM in crystals grown under similar conditions (approx. 6\lldens)\cite{liuTwoStepFluxSynthesis2023}. Fig. 3b also identfies the range of densities where a Wigner crystal is expected from theoretical estimates ($r_s>$ 30)\cite{drummondPhaseDiagramLowDensity2009}. We note that the critical density for the Wigner crystal is higher than the density estimated for the Anderson insulator, suggesting that formation of a Wigner crystal could be the origin of the MIT observed in our device. 
However, DC transport measurements alone cannot unambiguously differentiate between a Wigner crystal and localization due to disorder. Further studies including possibly optical\cite{smolenskiSignaturesWignerCrystal2021} or scanning probe measurements \cite{liImagingTwodimensionalGeneralized2021} willl be necesssary to resolve the orign of the low density mobility edge.

The robust contact achieved with our charge-transfer scheme also allows us to measure the properties of monolayer \wse under a strong magnetic field where interactions are significant due to the suppression of kinetic energy. Figure 4a shows the longitudinal resistance at 300 mK as a function of magnetic field and carrier density. We identify the presence of integer quantum Hall features down to $\nu=1$, and multiple fractional quantum Hall features in the $N=0$ and $N=1$ Landau levels (LLs). Integer and fractional quantum Hall states in monolayer \wse were previously identified in bulk compressibility measurements \cite{shiOddEvendenominatorFractional2020}, but their characteristic transport properties were inaccessible due to low performance contacts at low densities. Fig. 4b shows $R_{xx}$ and $1/R_{xy}$ as a function of filling fraction at 31.5 T and 300 mK. In the $N=0$ level, we observe minima or zeros in the longitudinal resistance at filling fractions $\nu=$2/3, 3/5, 1/3, 2/5, and 3/7. Additionally, we see the Hall conductance approaching the quantized values at filling fractions 2/3, 3/5, and 1/3. To our knowledge, this represents the first transport signature of fully developed FQH states in any van der Waals material beyond graphene. In the N=1 level, there are resistance minima at filling fractions $\nu=$ 6/5, 4/3, and 3/2 with quantized hall conductivity at $\nu=6/5$. 

The observed hierarchy of the FQH states in both the N=0 and N=1 LLs, including the appearance of an even denominator state in the N=1 level,  is in agreement with the hierarchy identified in bulk compressibility measurements\cite{shiOddEvendenominatorFractional2020}. 
In the N=0 LL we can estimate the gaps for the FQH states (Fig 4c,d) from the temperature dependence of the $R_{xx}$ minima. The gap magnitude appears to be approximately symmetric around $\nu=-1/2$ as expected for Laughlin states. Fitting to the expected scaling with filling fraction, $\Delta=\frac{eh}{m_{cyc}}B^{*}-\Gamma$ where $\Gamma$ is the disorder broadening and $B^{*}=B(1-2\nu)$\cite{polshynQuantitativeTransportMeasurements2018}, we find a disorder broadening of 8 K which is approximately consistent with the value identified from Shubnikov-de Haas oscillations (see SI). 
Further improvements in device quality will allow for a reduction in the disorder broadening and enable more thorough characterization of the fractional quantum Hall states. At $\nu=2/3$ the gap magnitude scales approximately as $\sqrt{B}$ as expected from the scaling of the effective composite fermion mass\cite{schulze-wischelerDirectMeasurementFactor2004}. The disorder broadening identified from this fit is approximately 9 K in agreement with the other measures of $\Gamma$.

Charge-transfer contacts to monolayer \wse have enabled the characterization of transport properties of a monolayer semiconductor with the highest mobility reported to date and access to densities an order of magnitude lower than previously reported. The improvement in sample and contact quality has revealed a MIT at a density where a Wigner crystal is expected to form and signatures of the fractional quantum Hall effect at high magnetic field. The access to transport properties of fractional quantum Hall states -- including an even denominator state -- in a monolayer semiconductor opens the door to new devices seeking to probe the nature of excitations of these correlated ground states. Moreover, the charge-transfer contact architecture presents new opportunities for engineering the contact interface for further improvements in contact performance. The ability to measure the transport properties of highly confined, high mobility carriers opens the door for new quantum devices where electronic transport can be strongly modified by mismatched material interfaces or artificial device patterning.

\section{Methods}

\subsection{Device fabrication}

Bulk \wse crystals are grown using a two-step flux synthesis in a 1:5 W:Se ratio\cite{liuTwoStepFluxSynthesis2023}. The monolayer \wse, BN dielectrics, and few-layer graphite flakes used to make the device are mechanically exfoliated from bulk crystals and identified by optical contrast. Prior to exfoliating \rucl, \sio substrates are coated with 1-dodecanol to enable later transfer steps. Silicon chips are placed onto a hot plate at 160 C and 1-dodecanol is applied to the surface. After 5 minutes, excess dodecanol is removed and chips are rinsed with isopropyl alcohol before drying with N$_2$. Flakes of \rucl are mechanically exfoliated onto this substrate and identified with optical contrast. 

The dual graphite gated monolayer \wse heterostructures are assembled from these flakes using the van der Waals dry transfer technique\cite{wangOneDimensionalElectricalContact2013} at temperatures ranging from 70-180 C using a polycarbonate transfer polymer. After assembly, the heterostructure is released onto a Silicon substrate with 285 nm of thermal oxide at 180 C.

After stacking, a metal contact gate is deposited on top of the heterostructure using electron beam lithography and electron beam evaporation. The metal electrodes to contact the gates and graphite contacts are first etched using a selective reactive ion etch using SF$_6$ and O$_2$ plasmas followed by metal deposition to make electrical contact. Finally, the device is etched into a Hall bar geometry using reactive ion etching with SF$_6$ and O$_2$ plasmas. During device fabrication, care is taken to avoid exposing the \rucl layer to solvents -- particularly acetone -- as this exposure can degrade the surface and reduce the magnitude of charge transfer. See supplementary material for additional fabrication details and device images. 

\subsection{Measurements}
Transport measurements were performed in a variable temperature cryostat with a base temperature of 1.5 K and in a 3He cryostat with a base temperature of 300 mK. Four and two terminal resistance measurements were carried out using a low-frequency lock-in technique at frequencies ranging from 3 Hz-200 Hz. IV characteristics were measured in DC using a sourcemeter. In all measurements, the bottom gate is grounded and the top gate is referred to as $V_G$.

\section{Acknowledgments}
\begin{acknowledgments}
This research is primarily supported by US Department of Energy (DE-SC0016703). Synthesis of WSe$_2$ (L.H., S.L., K.B.) was supported by the Columbia University Materials Science and Engineering Research Center (MRSEC), through NSF grants DMR-1420634 and DMR-2011738. A portion of this work was performed at the National High Magnetic Field Laboratory, which is supported by National Science Foundation Cooperative Agreement No. DMR-2128556 and the State of Florida. D.M. and M.C. acknowledge support from the Gordon and Betty Moore Foundation’s EPiQS Initiative, Grant GBMF9069. K.W. and T.T. acknowledge support from the JSPS KAKENHI (Grant Numbers 21H05233 and 23H02052) and World Premier International Research Center Initiative (WPI), MEXT, Japan.
\end{acknowledgments}
\section{Data availability}
The data that support the plots within this paper and other findings of this study are available from the corresponding author upon reasonable request.

\section*{Competing financial interests}
The authors declare no competing financial interests.

\bibliography{jp_refs_rucl3}

\end{document}